\documentclass{article}
\usepackage[a4paper, left=1.5cm, right=1.5cm, top=2.5cm, bottom=2.5cm]{geometry}
\usepackage{graphicx} % Required for inserting images
\usepackage{braket}
\usepackage{float}
\usepackage{amsmath}
\usepackage{amssymb}
\usepackage{xcolor}
\usepackage{authblk} % For author and affiliation formatting
\usepackage{cite}

\title{Smoking Gun Signatures of Quasilocal Probability in Black Hole Ringdowns}

\author[]{Oem Trivedi\thanks{Email: oem.trivedi@vanderbilt.edu}, Alfredo Gurrola\thanks{Email: alfredo.gurrola@vanderbilt.edu}, Robert J. Scherrer\thanks{Email: robert.scherrer@vanderbilt.edu} }

\affil[]{Department of Physics and Astronomy, Vanderbilt University, Nashville, TN 37235, USA}

\date{\today}

\begin{document}

\maketitle

\begin{abstract}
Building on recent work introducing the idea of Quasilocal Probability in curved spacetime, we develop its observational implications for black hole ringdown in detail. We show that horizon-induced probability flux leads to an effective non-Hermitian dynamics producing three distinctive signatures, which are correlated multi-mode deviations, weak amplitude dependence and a mismatch between waveform damping and energy accounting. These effects arise from a single boundary-flux mechanism and therefore exhibit a constrained, low-dimensional structure not expected in generic modified gravity scenarios. We demonstrate that while individual deviations may be mimicked, their combined pattern provides a robust discriminator of quasilocal probability. We further argue that upcoming gravitational wave observations can probe these signatures at meaningful precision. We  also establish that black hole ringdown is a novel arena to test whether quantum mechanical Hermiticity is really a fundamental property or an emergent symmetry in quantum gravity.
\end{abstract}

\section{Introduction}
Hermiticity is conventionally viewed as a foundational requirement of quantum mechanics within the Hilbert space formulation as the identification of physical observables with Hermitian operators ensures that measured quantities are real through spectral properties, enforces unitary time evolution in closed systems and underpins the probabilistic interpretation via the Born rule together with conservation of the standard inner product \cite{qm1bohm2013quantum,qm2zettili2009quantum,qm3sakurai2020modern,qm4griffiths2018introduction,qm5shankar2012principles,qm6scherrer2024quantum}. In this sense, Hermiticity is not simply a mathematical restriction but is deeply tied to measurement theory, probability conservation and the internal consistency of quantum dynamics. However, it has become increasingly evident that many realistic quantum systems are not perfectly isolated and instead they are effectively open due to coarse-graining, environmental interactions, continuous monitoring or fundamental inaccessibility of certain degrees of freedom \cite{nh1moiseyev2011non,nh2ashida2020non,nh3hatano1996localization}. In such contexts, effective descriptions naturally involve non-unitary evolution, complex potentials and gain-loss mechanisms. This has led to the systematic development of non-Hermitian quantum mechanics as a robust and experimentally accessible framework and modern realizations in controlled laboratory systems, where amplification, absorption and post-selection can be engineered, have enabled precise investigations of departures from strict unitarity. Consequently, non-Hermitian models are now treated both as effective descriptions of open systems and in some approaches, as generalized quantum theories in which consistency is maintained through modified inner products, metric structures and biorthogonal bases \cite{nh4jones2014relativistic,nh5gopalakrishnan2021entanglement,nh6hatano1997vortex,nh7bender2007making,nh8longhi2010optical}. This conceptual and experimental progress has further driven increasingly precise efforts to bound non-Hermitian effects using interferometric probes, spectral and pseudospectral diagnostics, decay measurements and explicit tests of norm conservation in appropriately generalized inner-product frameworks \cite{nh9jones2010non,nh10krejvcivrik2015pseudospectra,nh11cui2012geometric,nh12bergholtz2021exceptional}.
\\
\\
Recently we suggested a reinterpretation of Hermiticity as a symmetry principle \cite{hermTrivedi:2026qof} emerging from the conservation of the quantum inner product rather than as an a priori axiom. Within this perspective, when the inner-product current is globally conserved and no flux escapes through boundaries, the standard Hermiticity condition arises as the symmetry requirement on the generator of time evolution. Building on this viewpoint, a recent work \cite{qp1trivedi2026quasilocal} extended the framework to incorporate gravitational settings, where spacetime curvature and causal structure modify how probability conservation is implemented. In particular, horizons and causal boundaries can obstruct global conservation, leading naturally to a quasilocal formulation in which normalization and inner-product charge are defined relative to accessible regions and observers, which was termed as Quasilocal Probability (QP). This shift provides a physically motivated pathway toward understanding how effective non-Hermitian descriptions can arise in gravitational systems while remaining compatible with an underlying global conservation law.
\\
\\
While a preliminary discussion of observational signatures of QP was done in \cite{qp1trivedi2026quasilocal}, we take this direction a step further here and go into detail on observational signatures of this framework. In the next section we an slight overview of quasilocal probability and Hermiticity as a symmetry law. In section 3 we outline in detail the possible clear signatures of QP in Black Hole Ringdowns while in section 4 we detail how these signatures can be most naturally explained by QP and not by exotic/modified gravity constructions. In section 5 we talk about the observational feasibility of detecting such signatures and we conclude our work in section 6. 
\\
\\
\section{Hermiticity Symmetry and Quasilocal Probability}

The standard formulation of quantum mechanics is built upon the assumption that the Hamiltonian is Hermitian, ensuring real spectra, unitary time evolution and conservation of the canonical inner product \cite{qm1bohm2013quantum,qm2zettili2009quantum,qm3sakurai2020modern,qm4griffiths2018introduction,qm5shankar2012principles,qm6scherrer2024quantum}. However, a broader perspective emerges once one relaxes this requirement and allows for non-Hermitian generators while maintaining Schrodinger evolution,
\begin{equation}
i\hbar \frac{\partial}{\partial t}\ket{\psi(t)}=\hat H \ket{\psi(t)}
\label{eq:NH_Schro}
\end{equation}
in which case the Hamiltonian may always be decomposed as
\begin{equation}
\hat H=\hat H_{\rm H}+i\hat \Gamma
\label{eq:H_decomp}
\end{equation}
with $\hat H_{\rm H}$ Hermitian and $\hat \Gamma$ Hermitian as well. The presence of $\hat \Gamma$ modifies the conservation of probability, as the norm evolves according to
\begin{equation}
\frac{d}{dt}\braket{\psi(t)|\psi(t)} = \frac{i}{\hbar}\bra{\psi(t)}\big(\hat H^{\dagger}-\hat H\big)\ket{\psi(t)}
= -\frac{2}{\hbar}\bra{\psi(t)}\hat \Gamma\ket{\psi(t)}
\label{eq:norm_evol}
\end{equation}
so that amplification or decay is governed directly by the anti-Hermitian component and such effective descriptions naturally arise in open quantum systems, dissipative dynamics and coarse-grained settings \cite{nh1moiseyev2011non,nh2ashida2020non,nh3hatano1996localization}. The spectral structure is correspondingly generalized, requiring a biorthogonal basis
\begin{equation}
\hat H\ket{n_{\rm R}}=E_n\ket{n_{\rm R}}
\qquad
\hat H^{\dagger}\ket{n_{\rm L}}=E_n^{\ast}\ket{n_{\rm L}}
\qquad
\braket{n_{\rm L}|m_{\rm R}}=\delta_{nm}
\label{eq:biorth}
\end{equation}
which preserves consistency even for complex eigenvalues \cite{nh4jones2014relativistic,nh5gopalakrishnan2021entanglement,nh6hatano1997vortex,nh7bender2007making,nh8longhi2010optical}. In special cases, a modified inner product defined through a metric operator restores a generalized notion of unitarity, indicating that Hermiticity itself may be interpreted as a symmetry condition associated with conservation of a suitably defined inner product rather than a fundamental axiom.
\\
\\
This reinterpretation becomes more transparent when one considers inner-product conservation as the primary principle \cite{hermTrivedi:2026qof}. Demanding invariance of a generalized norm under time evolution leads to a condition of the form $\hat H^{\dagger}\eta=\eta \hat H$, which reduces to standard Hermiticity when $\eta=\mathbb{I}$. In relativistic field theory this idea acquires a covariant expression through the conserved current associated with the Klein-Gordon structure,
\begin{equation}
\nabla_{\mu}J^{\mu}=0
\end{equation}
with the corresponding inner-product charge defined on a hypersurface $\Sigma$ as
\begin{equation}
Q[\Sigma]=\int_{\Sigma} d\Sigma \, n_{\mu} J^{\mu}
\end{equation}
which is independent of $\Sigma$ in the absence of boundary flux. However, in curved spacetimes the situation is more subtle as restricting to a finite region $R$ introduces a boundary contribution, gives us
\begin{equation}
Q_{R}[\Sigma_2]-Q_{R}[\Sigma_1] = -\int_{B_{R}} d\Sigma_{\mu} J^{\mu}
\end{equation}
so that conservation is replaced by a flux balance law and this structure mirrors the behavior of energy in general relativity, where the Einstein equations imply
\begin{equation}
\nabla_{\mu}G^{\mu\nu}=0
\end{equation}
and lead to conserved currents such as
\begin{equation} \label{kill}
j^{\mu}_{(\xi)}=T^{\mu\nu}\xi_{\nu}
\end{equation}
whose associated charges become quasilocal in the presence of boundaries. In this sense  Hermiticity corresponds to the global realization of inner-product conservation, while gravitational causal structure determines whether such conservation can be implemented globally or only within restricted domains.
\\
\\
This naturally leads to a quasilocal formulation of probability \cite{qp1trivedi2026quasilocal}. For a region $R$, one defines
\begin{equation}
Q_{R}[\Sigma]=\int_{\Sigma\cap R} d\Sigma \, n_{\mu} J^{\mu}
\end{equation}
and the evolution of this quantity obeys the balance law
\begin{equation}
Q_{R}[\Sigma_{2}]-Q_{R}[\Sigma_{1}]
=
-\int_{B_{R}} d\Sigma_{\mu} J^{\mu}
\label{eq:quasilocal_balance}
\end{equation}
which provides a precise notion of quasilocal probability and when dynamics is restricted to $R$ and described by
\begin{equation} \label{evol}
i\hbar \frac{\partial}{\partial t}\Phi = H_{R}\Phi
\end{equation}
the change in the regional inner product satisfies
\begin{equation}
i(\Phi,(H_{R}-H_{R}^{\dagger})\Phi)_{R}
= -\int_{\partial R} dS \, J^{\mu}s_{\mu}
\label{bala}
\end{equation}
showing that the anti-Hermitian part of the effective generator is directly sourced by boundary flux. Thus, QP and effective non-Hermiticity are equivalent descriptions of the same physical mechanism. In globally closed systems or in flat spacetime with appropriate falloff the boundary term vanishes and standard Hermiticity is recovered. In contrast, in curved spacetimes with horizons or finite causal domains, such boundaries are unavoidable, and probability conservation becomes intrinsically quasilocal. This establishes a direct conceptual bridge between non-Hermitian quantum dynamics, gravitational causal structure and the emergence of observer-dependent notions of probability, extending the analogy with quasilocal energy in general relativity. 
\\
\\
\section{Quasinormal Probability Predictions for Ringdown}
We begin here from the fundamental observation that relativistic quantum fields admit a conserved inner-product current. For a complex scalar field, this current takes the form
\begin{equation}
J^\mu = -i\left(\Phi^\dagger \nabla^\mu \Phi - \Phi \nabla^\mu \Phi^\dagger\right)
\end{equation}
and satisfies the continuity equation
\begin{equation}
\nabla_\mu J^\mu = 0
\end{equation}
This equation implies that for any spacelike hypersurface $\Sigma$, the total inner-product charge
\begin{equation}
Q[\Sigma] = \int_\Sigma d\Sigma\, n_\mu J^\mu
\end{equation}
is conserved under evolution, provided no flux escapes through the boundary of the spacetime region under consideration.
\\
\\
However, in curved spacetimes containing horizons or causal boundaries, physically relevant observers are restricted to finite regions $R$. In such cases, one must instead consider the quasilocal charge defined by
\begin{equation}
Q_R[\Sigma] = \int_{\Sigma \cap R} d\Sigma\, n_\mu J^\mu
\end{equation}
Applying Gauss's theorem to the continuity equation over a spacetime volume bounded by two hypersurfaces $\Sigma_1$ and $\Sigma_2$ and the intervening boundary $B_R$, one obtains
\begin{equation}
\int_{\Sigma_2 \cap R} d\Sigma\, n_\mu J^\mu
- \int_{\Sigma_1 \cap R} d\Sigma\, n_\mu J^\mu
= - \int_{B_R} d\Sigma_\mu\, J^\mu
\end{equation}
which can be written as
\begin{equation}
Q_R[\Sigma_2] - Q_R[\Sigma_1] = - \int_{B_R} d\Sigma_\mu\, J^\mu
\end{equation}
Thus, the change in accessible probability within the region $R$ is entirely determined by the flux of the inner-product current across the boundary $B_R$. In particular, for black hole spacetimes, the event horizon contributes a null component to $B_R$, and ingoing modes generically lead to a nonvanishing flux into the horizon.
\\
\\
To construct an effective dynamical description for observers restricted to $R$, we introduce a regional inner product
\begin{equation}
(\Phi,\Psi)_R = \int_{\Sigma \cap R} d\Sigma\, n_\mu J^\mu(\Phi,\Psi)
\end{equation}
and assume that the field evolves according to an effective Schrodinger type equation \eqref{evol}. We now derive the condition that $H_R$ must satisfy in order to reproduce the quasilocal balance law as by taking the time derivative of the regional norm,
\begin{equation}
\frac{d}{dt}(\Phi,\Phi)_R = \langle \dot{\Phi} \mid \Phi \rangle_ + \langle \Phi \mid \dot{\Phi} \rangle_R
\end{equation}
Substituting in the evolution equation \eqref{evol},
\begin{equation}
\dot{\Phi} = -\frac{i}{\hbar} H_R \Phi,
\qquad \dot{\Phi}^\dagger = \frac{i}{\hbar} \Phi^\dagger H_R^\dagger
\end{equation}
we obtain
\begin{equation}
\frac{d}{dt}(\Phi,\Phi)_R = \frac{i}{\hbar} \langle \Phi \mid H_R^\dagger \mid \Phi \rangle_R - \frac{i}{\hbar} \langle \Phi \mid H_R \mid \Phi \rangle_R
\end{equation}
so that
\begin{equation}
\frac{d}{dt}(\Phi,\Phi)_R = \frac{i}{\hbar} \langle \Phi \mid (H_R^\dagger - H_R) \mid \Phi \rangle_R
\end{equation}
Comparing this with the flux expression,
\begin{equation}
\frac{d}{dt}(\Phi,\Phi)_R = - \int_{\partial R} dS\, J^\perp
\end{equation}
we conclude that
\begin{equation}
H_R - H_R^\dagger = -2i\Gamma
\end{equation}
where $\Gamma$ is a positive semi-definite operator encoding the boundary leakage of inner-product charge. This implies that the most general effective Hamiltonian consistent with the balance law is
\begin{equation}
H_R = H_0 - i\Gamma
\end{equation}
with $H_0^\dagger = H_0$ and $\Gamma^\dagger = \Gamma$.
We now specialize to perturbations around a black hole background and expand the field in a basis of quasinormal-mode-like functions,
\begin{equation}
\Phi(t,\mathbf{x}) = \sum_n a_n(t)\,\phi_n(\mathbf{x})
\end{equation}
where the basis functions satisfy appropriate boundary conditions at the horizon and infinity. We assume orthonormality with respect to the regional inner product,
\begin{equation}
(\phi_n,\phi_m)_R = \delta_{nm}
\end{equation}
Projecting the evolution equation onto this basis,
\begin{equation}
i\hbar \dot{a}_n = \sum_m \left[ (\phi_n,H_0\phi_m)_R - i(\phi_n,\Gamma\phi_m)_R \right] a_m
\end{equation}
Defining
\begin{equation}
H^{(0)}_{nm} = (\phi_n,H_0\phi_m)_R,
\qquad
\Gamma_{nm} = (\phi_n,\Gamma\phi_m)_R
\end{equation}
we obtain
\begin{equation}
i\hbar \dot{a}_n = \sum_m \left( H^{(0)}_{nm} - i\Gamma_{nm} \right) a_m
\end{equation}
Diagonalizing the Hermitian part,
\begin{equation}
H^{(0)}_{nm} = \hbar \omega_n^{(0)} \delta_{nm}
\end{equation}
the evolution equation becomes
\begin{equation}
i\hbar \dot{a}_n = \hbar \omega_n^{(0)} a_n - i \sum_m \Gamma_{nm} a_m
\end{equation}
This provides the complete dynamical system from which all observational signatures follow. \\

\subsection{Correlated multi-mode deviations}
We first consider the case in which off-diagonal elements of $\Gamma_{nm}$ are small compared to diagonal ones. Then
\begin{equation}
i\hbar \dot{a}_n = \hbar \omega_n^{(0)} a_n - i \Gamma_{nn} a_n
\end{equation}
so that
\begin{equation}
\dot{a}_n = -i\omega_n^{(0)} a_n - \frac{\Gamma_{nn}}{\hbar} a_n
\end{equation}
Integrating this equation yields
\begin{equation}
a_n(t) = a_n(0)\exp\left[-i\omega_n^{(0)} t - \frac{\Gamma_{nn}}{\hbar} t \right]
\end{equation}
and therefore the corrected complex frequency is
\begin{equation}
\omega_n = \omega_n^{(0)} - i\frac{\Gamma_{nn}}{\hbar}
\end{equation}
The crucial feature is that $\Gamma_{nn}$ arises from the same boundary flux operator for all modes. Writing explicitly,
\begin{equation}
\Gamma_{nn} = \int_{\partial R} dS\, \mathcal{F}_{nn}
\end{equation}
where $\mathcal{F}_{nn}$ depends on the mode profile at the boundary, one can factorize
\begin{equation}
\mathcal{F}_{nn} = \lambda W_n
\end{equation}
with $\lambda$ a universal leakage parameter determined by the geometry and $W_n$ a mode-dependent weight. Then
\begin{equation}
\delta \omega_n = -i \lambda W_n
\end{equation}
and thus
\begin{equation}
\frac{\delta \omega_n}{W_n} = \frac{\delta \omega_m}{W_m}
\end{equation}
for all modes $n,m$, demonstrating that deviations are not independent but correlated and now including off-diagonal terms, the system becomes
\begin{equation}
i\hbar \dot{\mathbf{a}} = (\Omega^{(0)} - i\Gamma)\mathbf{a}
\end{equation}
and the eigenvalue problem gives
\begin{equation}
\omega_n = \omega_n^{(0)} - \frac{i}{\hbar}\Gamma_{nn} - \frac{1}{\hbar^2}\sum_{m \neq n} \frac{\Gamma_{nm}\Gamma_{mn}}{\omega_n^{(0)} - \omega_m^{(0)}}
\end{equation}
which again shows that all corrections are constructed from the same matrix $\Gamma_{nm}$.
\begin{figure}[!h]
\centering
\includegraphics[width=1\linewidth]{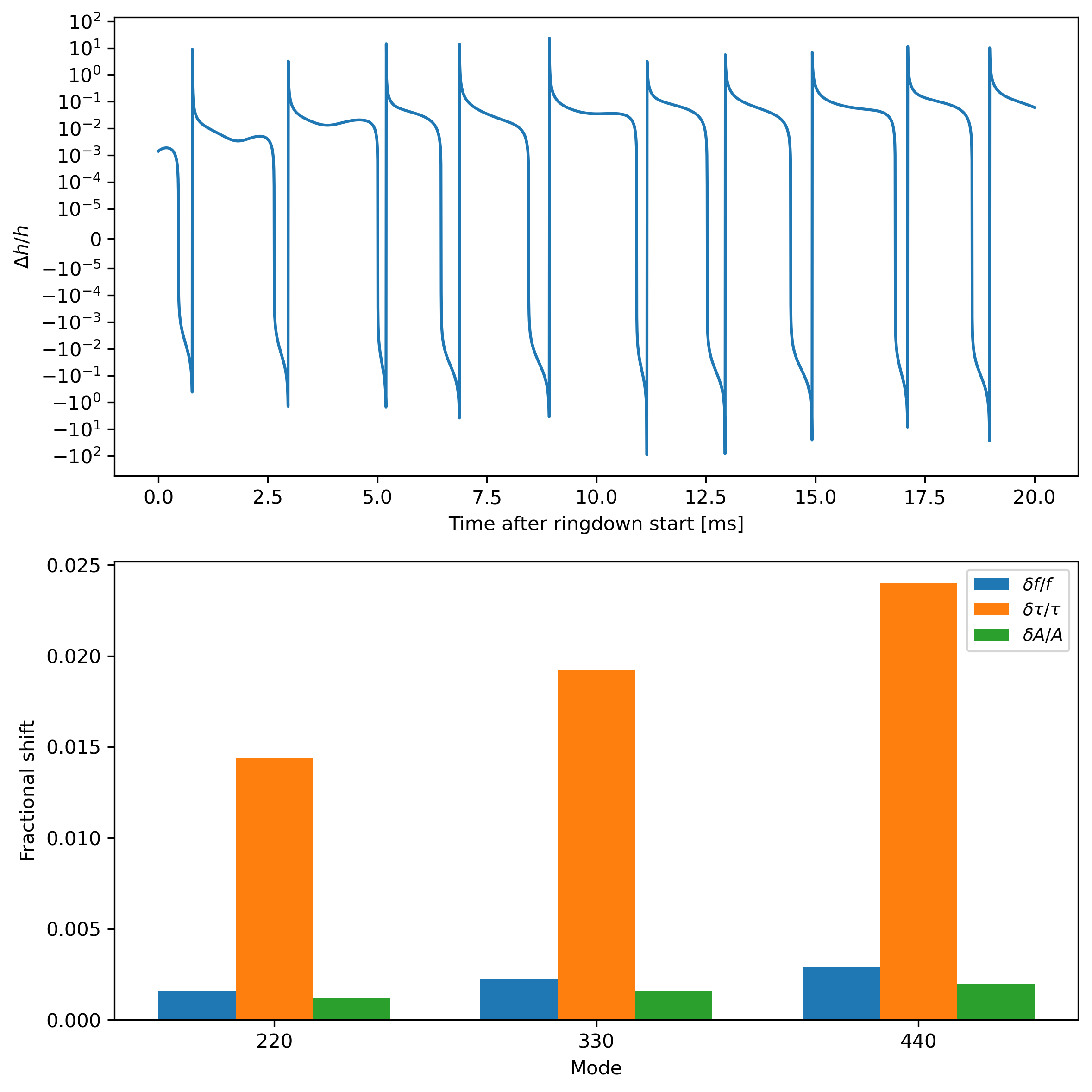}
\caption{Illustrative multimode ringdown deviations in the quasilocal probability framework. The top panel shows the relative residual $\Delta h/h$ between the quasilocal waveform and the Kerr prediction, plotted on a symmetric logarithmic scale. The residual exhibits a structured oscillatory pattern with sharp features arising from correlated shifts in multiple quasinormal modes, with enhanced excursions near waveform nodes. The bottom panel displays the fractional deviations in frequency, damping time, and amplitude for the $(\ell,m,n)=(220,330,440)$ modes. All deviations are controlled by a single quasilocal parameter $\epsilon$, demonstrating the correlated multi-mode structure predicted by the framework.}
\label{multimode}
\end{figure}

The multimode structure of the QP corrections follows directly from the effective evolution equation derived earlier, where the regional Hamiltonian takes the form $H_R = H_0 - i\Gamma$. Expanding the field in a quasinormal mode basis, the amplitudes satisfy a coupled system in which both diagonal and off-diagonal elements of the operator $\Gamma_{nm}$ are determined by boundary flux of the inner-product current. Since this flux originates from the same geometric boundary for all modes, the resulting corrections to the complex frequencies are not independent. Instead, each mode acquires a shift proportional to a common underlying parameter, leading to the schematic structure $\delta \omega_n \propto \Gamma_{nn}$, with all $\Gamma_{nn}$ derived from the same operator. This immediately implies that the fractional deviations in frequency and damping time across different modes are correlated rather than arbitrary.
\\
\\
At the waveform level, this correlated structure manifests as a coherent but highly structured deviation in the relative amplitude of the multimode signal. When expressed as a fractional residual $\Delta h/h$, the deformation reveals a distinctive oscillatory pattern with sharp, regularly spaced features as seen in the top panel of figure \ref{multimode}. These features arise from the interference between slightly shifted quasinormal modes and are enhanced near the zero crossings of the waveform, making the correlated nature of the deformation particularly transparent. The logarithmic scaling further shows the multi-scale structure of the deviation, allowing both small and large excursions to be visualized simultaneously. The bottom panel illustrates that the fractional shifts $\delta f_{\ell mn}/f_{\ell mn}$, $\delta \tau_{\ell mn}/\tau_{\ell mn}$ and $\delta A_{\ell mn}/A_{\ell mn}$ remain proportional to the same small parameter $\epsilon$, confirming that the entire pattern is governed by a single underlying physical mechanism. This highlights the defining feature of the quasilocal probability framework, namely that it predicts a structured, low dimensional deformation of the Kerr spectrum rather than independent phenomenological shifts of individual modes.\\

\subsection{Amplitude and state dependence}
To capture possible nonlinear effects, we allow the boundary flux to depend on the field amplitude and expanding,
\begin{equation}
J^\perp = J^{\perp}_{(0)} + \alpha |\Phi|^2 J^{\perp}_{(1)}
\end{equation}
this induces
\begin{equation}
\Gamma = \Gamma^{(0)} + \alpha \Gamma^{(1)}[\Phi]
\end{equation}
Projecting onto a single dominant mode,
\begin{equation}
i\hbar \dot{a}_n = \hbar \omega_n^{(0)} a_n - i \Gamma_n(|a_n|^2) a_n
\end{equation}
with
\begin{equation}
\Gamma_n(|a_n|^2) = \Gamma_n^{(0)} + \alpha_n |a_n|^2
\end{equation}
Thus
\begin{equation}
\dot{a}_n = -i\omega_n^{(0)} a_n - \frac{1}{\hbar} \left(\Gamma_n^{(0)} + \alpha_n |a_n|^2 \right) a_n
\end{equation}
Writing $a_n = A_n e^{-i\varphi_n}$,
\begin{equation}
\dot{A}_n = -\frac{1}{\hbar} \left(\Gamma_n^{(0)} + \alpha_n A_n^2 \right) A_n
\end{equation}
which leads to
\begin{equation}
\frac{d}{dt}(A_n^2) = -\frac{2\Gamma_n^{(0)}}{\hbar} A_n^2 - \frac{2\alpha_n}{\hbar} A_n^4
\end{equation}
Hence the damping rate depends on amplitude,
\begin{equation}
\gamma_{\text{eff}}(A_n) = \frac{\Gamma_n^{(0)}}{\hbar} + \frac{\alpha_n}{\hbar} A_n^2
\end{equation}
\begin{figure}[!h]
    \centering
    \includegraphics[width=1\linewidth]{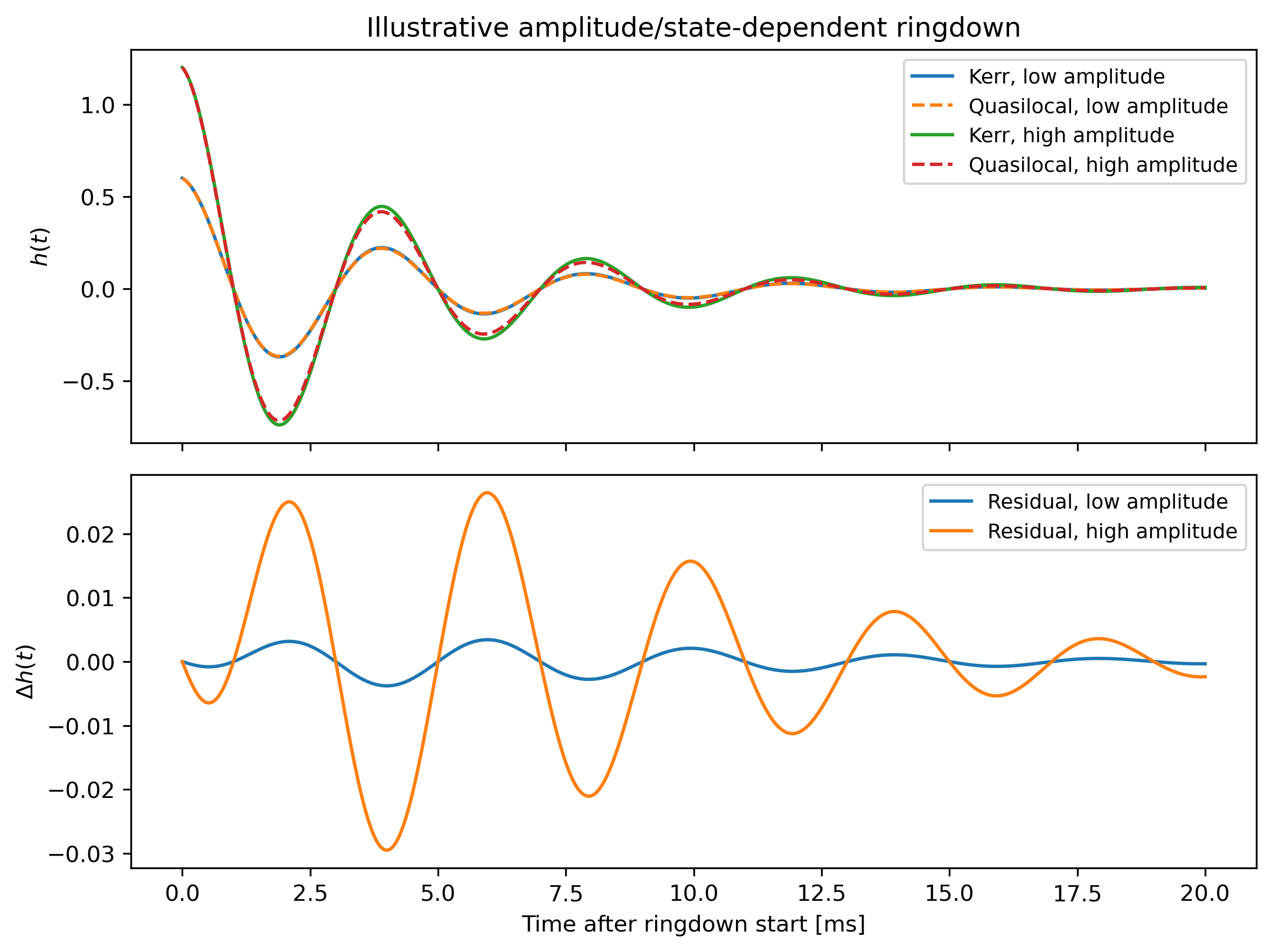}
    \caption{Illustrative amplitude-dependent ringdown in the quasilocal probability framework. The top panel compares Kerr and quasilocal waveforms for two different initial amplitudes. While the Kerr evolution is independent of amplitude apart from an overall scaling, the quasilocal waveform exhibits a slight amplitude dependence, with larger initial excitations deviating more strongly. The bottom panel shows the residual $\Delta h(t)=h_{\rm QL}(t)-h_{\rm Kerr}(t)$ for both cases. The higher-amplitude state produces a significantly larger residual, demonstrating that the effective damping depends on the state of the system. This behavior arises from the amplitude dependence of the boundary flux and constitutes a distinctive nonlinear signature of QP.}
    \label{ampring}
\end{figure}
The amplitude dependence of the ringdown signal follows from the generalized form of the effective Hamiltonian, where the anti-Hermitian operator $\Gamma$ inherits a functional dependence on the field configuration through the boundary flux. In this case, the flux of the inner product current is no longer strictly linear in the field amplitude, but instead admits an expansion in powers of $|\Phi|^2$. Upon projection onto a dominant quasinormal mode, this leads to an effective damping term that depends on the instantaneous amplitude of the mode. As a result, the evolution equation for the mode amplitude acquires a nonlinear contribution such that the decay rate is no longer constant but depends on the excitation level.
\\
\\
At the waveform level, this comes as a state dependent modification of the ringdown signal and for small amplitudes, the nonlinear contribution is negligible and the evolution closely tracks the Kerr prediction. However, for larger excitations, the additional damping term becomes more significant leading to a visibly altered decay envelope. This is reflected in the residuals shown in the lower panel of Fig. \ref{ampring}, where the higher amplitude case exhibits a substantially larger deviation from Kerr. The key point is that this effect cannot be mimicked by a simple rescaling of the waveform, as it arises from a genuine dependence of the effective dynamics on the state itself. This provides a direct observational handle on the underlying QP flux. \\

\subsection{Mismatch with energy accounting}
Energy is governed by the conserved current
\begin{equation}
j^\mu = T^{\mu\nu} \xi_\nu, \qquad \nabla_\mu j^\mu = 0
\end{equation}
with quasilocal energy
\begin{equation}
E_R = \int_{\Sigma \cap R} d\Sigma\, n_\mu j^\mu
\end{equation}
and balance law
\begin{equation}
\frac{dE_R}{dt} = - \int_{\partial R} dS\, j^\perp
\end{equation}
Meanwhile we have
\begin{equation}
Q_n = |a_n|^2
\end{equation}
satisfies the relation
\begin{equation}
\dot{Q}_n = -\frac{2\Gamma_n}{\hbar} Q_n
\end{equation}
so that
\begin{equation}
Q_n(t) = Q_n(0) e^{-2\gamma_n t},
\qquad
\gamma_n = \frac{\Gamma_n}{\hbar}
\end{equation}
Energy evolves as
\begin{equation}
\frac{dE_n}{dt} = -2\kappa_n E_n
\end{equation}
with $\kappa_n$ determined by energy flux and since $\Gamma_n$ is determined by inner product flux rather than energy flux, one generally has
\begin{equation}
\gamma_n = \kappa_n + \varepsilon_n
\end{equation}
and therefore
\begin{equation}
\frac{\tau_n^{\text{obs}} - \tau_n^{\text{energy}}}{\tau_n^{\text{energy}}}
= -\frac{\varepsilon_n}{\kappa_n}
\end{equation}
which quantifies the mismatch between damping inferred from waveform decay and that inferred from energy conservation.
\begin{figure}[!h]
\centering
\includegraphics[width=1\linewidth]{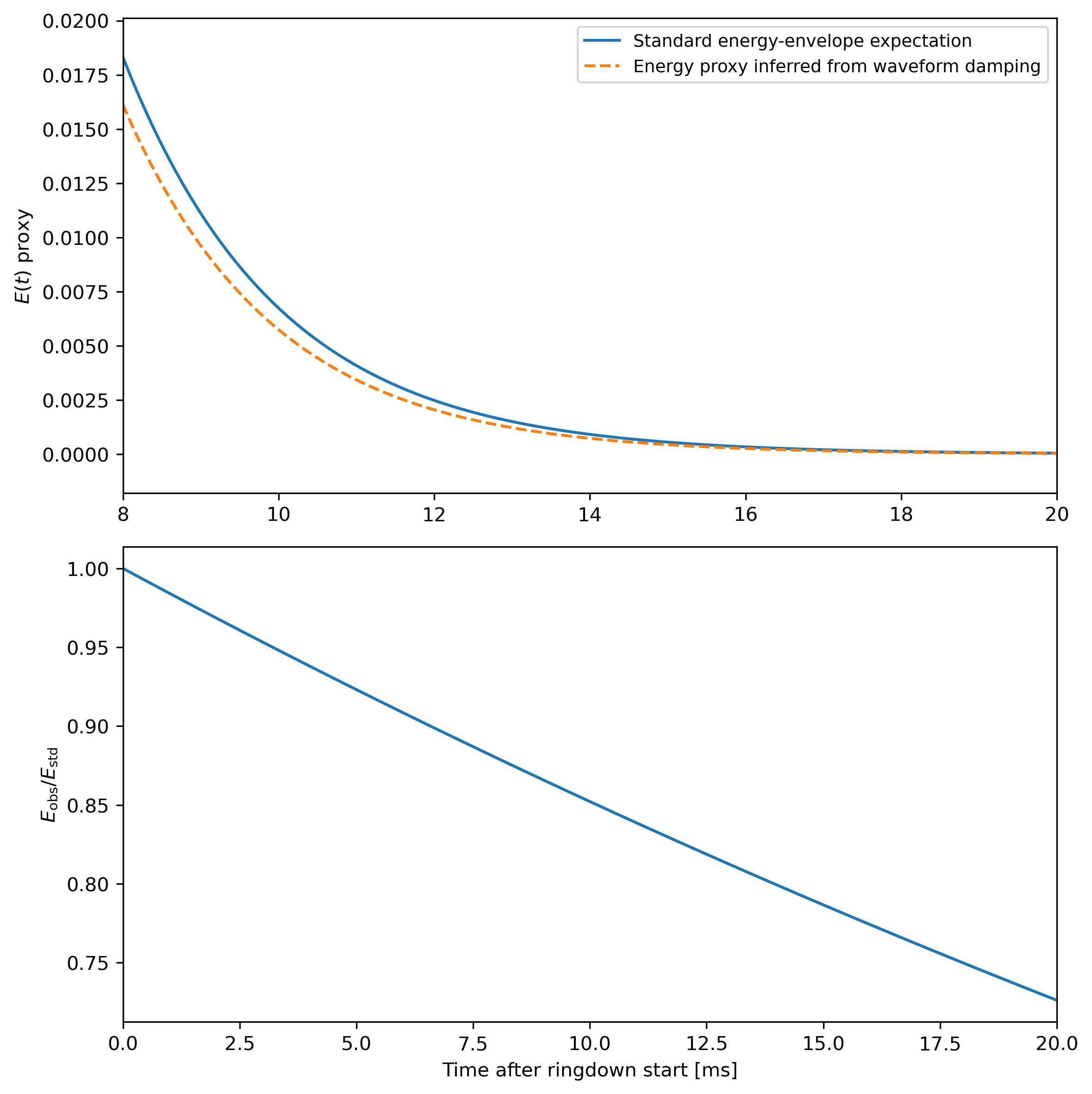}
\caption{Illustrative mismatch between waveform damping and energy-envelope expectations. The top panel compares the standard exponential energy decay (solid blue) with the energy proxy inferred from the observed waveform damping (dashed orange), showing that the two remain very even at the level of the absolute signal. The bottom panel displays the ratio $E_{\rm obs}/E_{\rm std}$, which reveals a systematic deviation from unity due to the presence of an additional damping contribution. This relative measure makes explicit the cumulative discrepancy between the observed decay and the standard energy-accounting prediction.}
\label{energymismatch}
\end{figure}
The mismatch between waveform damping and energy accounting arises from the fact that the effective non-Hermitian contribution to the regional Hamiltonian is governed by the flux of the inner-product current, whereas energy evolution is controlled by the flux of the stress energy tensor. Although both quantities obey continuity equations, they correspond to distinct conserved currents and as a result, the decay rate inferred from the waveform amplitude, which is directly related to the loss of QP, need not coincide exactly with the decay rate expected from energy flux considerations.
\\
\\
At the level of a single quasinormal mode the waveform amplitude decays as $\exp(-\gamma t)$, implying an energy like proxy proportional to $\exp(-2\gamma t)$. In contrast, the standard energy balance would predict a decay governed by a rate $\kappa$, leading to $\exp(-2\kappa t)$ and if $\gamma \neq \kappa$, the ratio of these two quantities evolves as $\exp[-2(\gamma-\kappa)t]$, producing a cumulative deviation over time. This behavior is illustrated in Fig. \ref{energring}, where the waveform itself shows small deviations, but the derived energy proxy reveals a systematic mismatch. This provides a direct observational handle on the distinction between probability flux and energy flux in the quasilocal framework. \\

\section{Discriminating Quasilocal Probability from Alternative Explanations}

\begin{figure}[!h]
    \centering
    \includegraphics[width=0.9\linewidth]{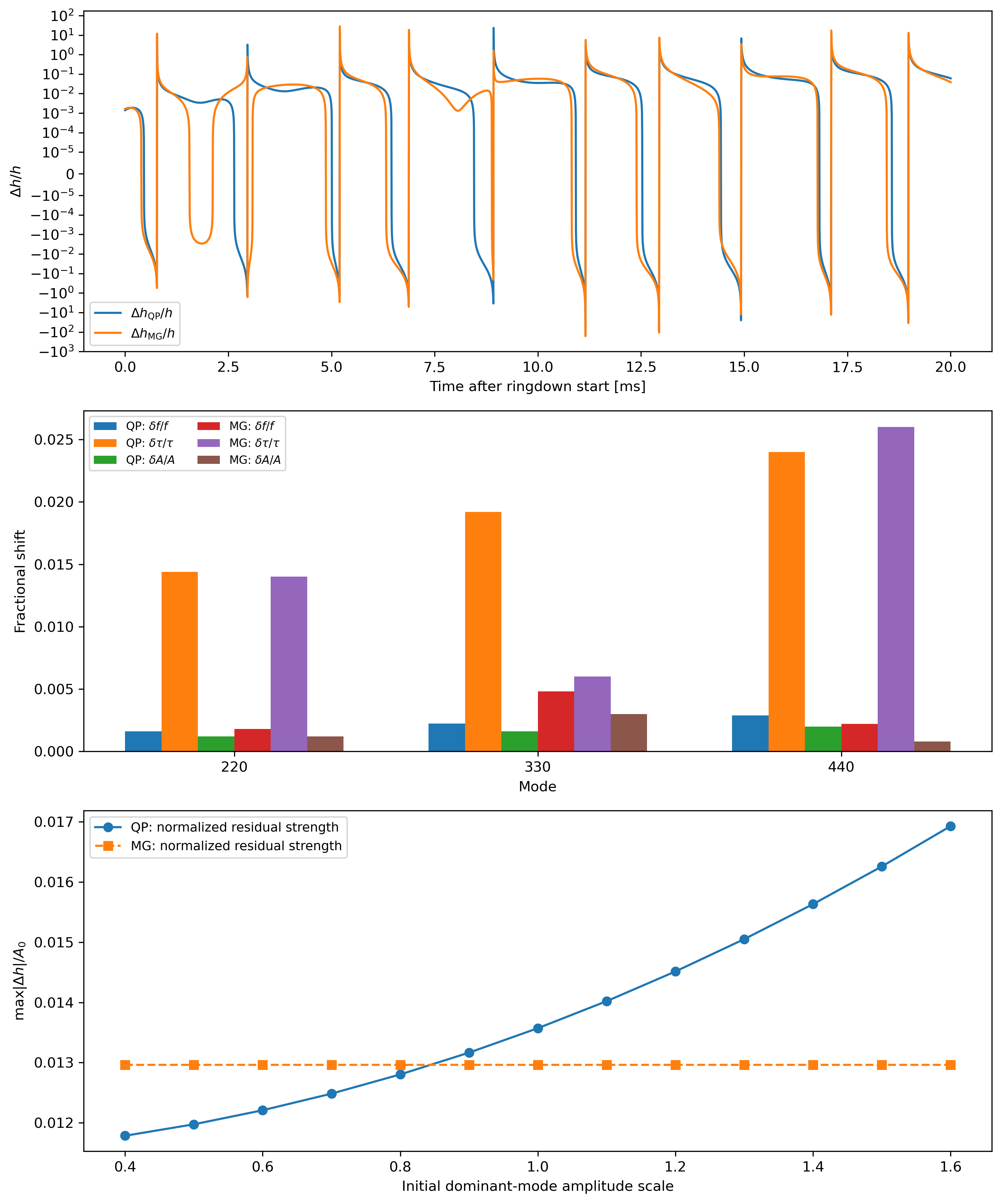}
    \caption{Illustrative comparison between quasilocal probability and generic modified gravity deformations of the multimode ringdown. The top panel shows the relative residual $\Delta h/h$ for both scenarios on a symmetric logarithmic scale, wherein we see that in the quasilocal case (blue) a coherent oscillatory structure arises from correlated multimode shifts, while the modified gravity case (orange) displays a more irregular pattern due to independent parameter variations. The middle panel compares the fractional deviations in frequency, damping time and amplitude for the $(\ell,m,n)=(220,330,440)$ modes. In the quasilocal framework, all shifts are controlled by a single parameter, whereas in the modified gravity case they are independent and mode dependent. The bottom panel shows the normalized residual strength as a function of the initial dominant-mode amplitude.}
    \label{qpvsmg}
\end{figure}

A central question is whether the observational signatures derived above can be uniquely attributed to QP effects, or whether they could be mimicked by generic modifications of gravity or other exotic near-horizon physics. For investigating this we begin again from the effective evolution equation for the restricted system
\begin{equation*}
i\hbar \frac{d}{dt}\mathbf{a} = \left(\Omega^{(0)} - i\Gamma\right)\mathbf{a}
\end{equation*}
Note that the the crucial point is that $\Gamma$ is not an arbitrary matrix, but is constructed from a single geometric source, namely the flux integral over the boundary,
\begin{equation*}
\Gamma_{nm} = \int_{\partial R} dS\, \mathcal{F}_{nm}[\phi_n,\phi_m]
\end{equation*}
This implies that all corrections to the mode spectrum are governed by a single operator and in particular, the first order shifts in the complex frequencies are given by
\begin{equation}
\delta \omega_n = -\frac{i}{\hbar} \Gamma_{nn}
\end{equation}
and therefore the fractional shifts satisfy
\begin{equation}
\frac{\delta \omega_n}{\omega_n^{(0)}} = -\frac{i}{\hbar}\frac{\Gamma_{nn}}{\omega_n^{(0)}}
\end{equation}
Since each $\Gamma_{nn}$ arises from the same boundary functional, one may write
\begin{equation}
\Gamma_{nn} = \lambda W_n
\end{equation}
with $\lambda$ a universal parameter taking into account the strength of QP leakage and $W_n$ determined by the mode profile and so we have
\begin{equation}
\delta \omega_n = -i \lambda W_n
\end{equation}
which immediately implies that all mode deviations lie on a one-parameter family. This low dimensional structure is a direct consequence of the flux origin of $\Gamma$ but in contrast, in generic modified gravity theories deviations in quasinormal modes arise from modifications of the background metric or perturbation potential. If the effective radial equation takes the form
\begin{equation}
\frac{d^2\psi}{dr_*^2} + \left[\omega^2 - V(r)\right]\psi = 0
\end{equation}
then a modification $V \to V + \delta V$ leads to shifts
\begin{equation}
\delta \omega_n = \langle \psi_n \mid \delta V \mid \psi_n \rangle
\end{equation}
where the inner product is defined over the radial domain. In this case, $\delta V(r)$ is an arbitrary function and the resulting shifts $\delta \omega_n$ are independent for different modes unless additional structure is imposed and so we see that generic modifications of the potential lead to a high-dimensional parameter space of deviations,
\begin{equation}
\left\{ \delta \omega_n \right\} \sim \text{independent for each } n
\end{equation}
in contrast to the one parameter structure emerging from quasilocal probability.
\\
\\
A second distinguishing feature arises from the amplitude dependence discussed earlier wherein the quasilocal framework, the boundary flux admits an expansion in powers of the field amplitude,
\begin{equation}
\Gamma = \Gamma^{(0)} + \alpha \Gamma^{(1)}[\Phi]
\end{equation}
leading to an effective damping rate
\begin{equation}
\gamma_n(A) = \frac{\Gamma_n^{(0)}}{\hbar} + \frac{\alpha_n}{\hbar} A^2
\end{equation}
This dependence follows directly from the nonlinear structure of the flux, which is quadratic in the field and therefore naturally generates higher-order contributions. In linear perturbation theory around modified gravity backgrounds, the evolution equation remains linear in the perturbation
\begin{equation}
\frac{d^2\psi}{dt^2} + \mathcal{L}\psi = 0
\end{equation}
and therefore the damping rate is independent of the amplitude. Any amplitude dependence in such theories would require explicit nonlinear interactions beyond the linearized regime, which are typically suppressed and not directly tied to horizon structure. Thus, the presence of systematic amplitude dependence in the ringdown signal points naturally toward a boundary flux origin.
\\
\\
The third distinguishing feature concerns the mismatch between probability leakage and energy flux as in the quasilocal framework, probability is governed by the inner-product current $J^\mu$ while energy is governed by the stress-energy current $j^\mu$ and these currents satisfy independent conservation laws,
\begin{equation}
\nabla_\mu J^\mu = 0, \qquad \nabla_\mu j^\mu = 0
\end{equation}
but their associated fluxes across the boundary need not coincide. As a result, the decay rate of the waveform amplitude, controlled by $\Gamma$ can differ from the decay rate inferred from energy loss as discussed before but in standard and modified gravity theories, energy loss is typically the sole mechanism governing waveform damping. The energy carried by gravitational waves is directly related to the amplitude of the perturbation, and so
\begin{equation}
E(t) \propto h(t)^2 \propto e^{-2\kappa t}
\end{equation}
The damping rate inferred from the waveform is consistent with the energy flux in this. 
\\
\\
At the waveform level, the distinction between quasilocal probability and generic exotic deformations becomes apparent in the structure of the relative residual. When expressed as $\Delta h/h$, the quasilocal case produces a coherent oscillatory pattern with regular, correlated features across the entire ringdown while the modified gravity scenario leads to a more irregular deformation, as seen in the top panel of figure \ref{qpvsmg}. In the quasilocal case, the residual retains a clear multimode interference structure, reflecting the fact that all deviations are tied to a single underlying parameter. By contrast, the modified gravity residual exhibits a less organized pattern, arising from independent shifts in frequency, damping, and amplitude that are not constrained by a common physical origin. The middle panel further illustrates this distinction by comparing the fractional shifts in mode parameters. In the quasilocal framework, all deviations scale proportionally with the same small parameter, whereas in the modified gravity case the shifts are mode dependent and uncorrelated. This difference becomes even more pronounced when considering the response to changes in the initial state. The bottom panel shows that the normalized residual strength in the quasilocal case varies systematically with the initial dominant mode amplitude, showing an intrinsic state dependence, while the modified gravity scenario remains essentially insensitive to such changes. These features highlight that the quasilocal probability framework predicts a structured, correlated, and state dependent deformation of the ringdown signal, which is in stark contrast to the more arbitrary modifications expected from generic phenomenological models.
\section{Observational Feasibility of Signature Detections}

The feasibility of detecting the signatures discussed above is ultimately determined by the current precision of ringdown measurements and the projected sensitivity of future gravitational wave detectors. At present, analyses of LIGO–Virgo–KAGRA (LVK) data constrain the fractional deviations in the dominant quasinormal mode frequency at the $\sim 10\%$ level for individual events, with somewhat weaker constraints on damping times, often at the $\sim 10\%$-$40\%$ level depending on the signal-to-noise ratio and modelling assumptions \cite{ob1LIGOScientific:2025wao,ob2Ghosh:2021mrv}. Even in combined analyses, uncertainties remain at the several percent level for the best constrained quantities and this implies that any QP effect entering at the level $\epsilon \lesssim 10^{-2}$ would currently be difficult to detect directly through single-mode deviations alone. However, the key point is that the signatures proposed here are not isolated deviations but structured patterns across multiple observables, which can enhance detectability through correlated inference.
\\
\\
A particularly promising avenue is black hole spectroscopy, where multiple quasinormal modes are measured simultaneously . Current observations are only beginning to probe subdominant modes and in most events only the dominant $(220)$ mode is measured with high confidence. Nevertheless, theoretical and data analysis studies indicate that the observation of multiple modes could soon become routine, enabling precision tests of the Kerr spectrum \cite{sup1brito2015superradiance,sup2roy2022superradiance,sup3huang2019superradiant}. Importantly, higher overtones and subdominant modes are expected to provide stronger constraints on deviations from general relativity, as their frequencies and damping times are more sensitive to perturbations of the underlying dynamics. In this context, the correlated multi-mode structure predicted by QP is particularly advantageous, since even modest individual uncertainties can combine to constrain a shared parameter $\epsilon$ more tightly than independent mode-by-mode analyses.
\\
\\
Looking ahead, next generation detectors such as Cosmic Explorer \cite{ce1reitze2019cosmic,ce2evans2021horizon,ce3hall2022cosmic}, Einstein Telescope \cite{et1:2025xjr,et2Branchesi:2023mws,et3:2019dnz} and space based observatories like LISA \cite{lisa1amaro2017laser,lisa2auclair2023cosmology,lisa3amaro2023astrophysics,lisa4danzmann1996lisa} are expected to significantly improve ringdown sensitivity. Forecasts suggest that percent level or even subpercent precision on quasinormal mode frequencies may be achievable for high signal-to-noise events, and that nonlinear or secondary modes could become detectable in favorable cases. This improvement in precision directly translates into sensitivity to smaller values of $\epsilon$, potentially reaching the regime $\epsilon \sim 10^{-2}$ or below, where the quasilocal signatures discussed here would become observationally testable.
\\
\\
It is also important to emphasize that the detectability of quasilocal probability effects does not rely solely on resolving small absolute deviations, but on identifying their internal consistency structure. The correlated multi-mode deviations, weak amplitude dependence and possible mismatch between waveform damping and energy accounting together define a low-dimensional signature space that can be probed statistically across multiple events. Even if individual events do not achieve high precision, stacking analyses and hierarchical inference can reduce uncertainties and test whether deviations obey the predicted correlations. In this sense, the observational prospects are more favorable than a naive estimate based on single parameter sensitivity might suggest and the upcoming era of precision gravitational wave astronomy provides a realistic pathway toward testing the quasilocal probability framework very robustly and optimistically.
\\
\\
\section{Conclusions}
In this work, we have proposed a set of observational signatures arising from the quasilocal probability framework applied to black hole ringdown and shown how they provide a distinctive and testable departure from the standard Kerr paradigm. The three key signatures are namely a correlated multi-mode deviations, weak amplitude/state dependence and a mismatch between waveform damping and energy accounting. They all emerge from a common underlying principle, that pertaining to the flux of inner-product charge across a causal boundary. Unlike generic modifications of the gravitational background or phenomenological deformations of the perturbation potential, these effects are not independent and instead, they are tied together through a single effective operator governing probability leakage, leading to a constrained, low-dimensional structure across observables. We have demonstrated that while individual waveform deviations may be mimicked by suitably tuned exotic models, the simultaneous presence of all three signatures is highly nontrivial to reproduce without invoking a boundary-flux origin. Moreover, given the rapid progress in gravitational wave observations, particularly in multimode ringdown measurements and precision spectroscopy, these signatures lie within the realm of observational feasibility in the near future especially with the advent of next-generation detectors.
\\
\\
The implications of these results extend beyond phenomenology and touch upon foundational aspects of quantum theory in curved spacetime. In particular, the QP framework provides a concrete setting in which Hermiticity, traditionally treated as a fundamental postulate, can instead be viewed as an emergent symmetry that may be locally obstructed by causal structure. The presence of effective non-Hermitian terms, arising from inaccessible flux across horizons, leads directly to the signatures identified in this work. Therefore, observational verification of these effects would constitute evidence that Hermiticity is not an absolute property of the fundamental theory, but rather a symmetry that holds only under global conditions where probability is fully accessible. Conversely, increasingly stringent null results would place quantitative bounds on deviations from Hermiticity in gravitational settings. In this sense, black hole ringdown provides a unique laboratory for probing one of the most basic structural assumptions of quantum mechanics itself, opening the possibility of testing whether Hermiticity itself is a symmetry of nature.
\\
\\
\section*{Acknowledgements}
We gratefully acknowledge support from Vanderbilt University and the U.S. National Science Foundation. The work of OT is supported in part by the Vanderbilt Discovery Doctoral Fellowship. The work of AG is supported in part by NSF Award PHY-2411502. OT also thanks Florencia Aravena (Floppy) for many insightful conversations and support.

\bibliography{references}
\bibliographystyle{unsrt}

\end{document}